# Comparison of IQMD results with experimental data for Sn induced reactions

Bahadur Singh[1],* Suneel Kumar[1] and Rajeev K. Puri[2]
[1]School of Physics and Materials Science, Thapar University Patiala – 147004, INDIA and
[2]Department of Physics, Panjab University, Chandigarh – 160014 INDIA
* email: suneel.kumar@thapar.edu

## Introduction

In present days the most prominent part of nuclear physics is to study the properties of nuclear matter by intermediate heavy ion collision. It allows searching for a large number of observables which can be used to predict some conclusions about the nuclear matter under high densities and excitation energies [1]. The production of intermediate mass fragments (IMF's) has dependence on such observables one of them is $Z_{bound}$ (sum of all atomic number $Z_i$ of all projectile fragments with $Z_i>2$).

We have simulated three reactions $_{57}La^{124}+_{50}Sn^{124}$, $_{50}Sn^{124}+_{50}Sn^{124}$ and $_{50}Sn^{107}+_{50}Sn^{124}$ at energy 600 MeV/nucleon at different scaled impact parameters. The simulation has been carried out in the framework of Isospin dependent Quantum Molecular Dynamics (IQMD) model [2]. The isospin treated explicitly IQMD model by including Coulomb potential and different NN cross sections and symmetry potential.

## The Model: IQMD

Semi classical microscopic *improved version* of QMD model which is based on event by event method & includes.

1) **Isospin** dependent Coulomb potential.
2) Symmetry Potential.
3) NN cross-section.

Heavy ion collisions are simulated by generating the phase space (x, y, z, $p_x$, $p_y$, $p_z$) of two colliding nuclei at different time steps such that (x, y, z) are the position coordinates and ($p_x$, $p_y$, $p_z$) are the momentum coordinates.
The three steps of simulation are:

1) Initialization of projectile and target.

$$f_i(\vec{r},\vec{p},t) = \frac{1}{(\pi\hbar)^3} \times e^{[-(\vec{r}-\vec{r}_i(t))^2 \frac{2}{L}]} \times e^{[-(\vec{p}-\vec{p}_i(t))^2 L]}$$

2) Propagation of ($A_T+A_P$) nucleon system:

Each nucleon propagates under the classical Hamilton's equations of motion, given by

$$\frac{dr_i}{dt} = \frac{d\langle H\rangle}{dp_i}, \quad \frac{dp_i}{dt} = -\frac{d\langle H\rangle}{dr_i}$$

Where $\langle H\rangle = \langle T\rangle + \langle V\rangle$ is the Hamiltonian.

$V^{ij} = V^{ij}_{skyrme} + V^{ij}_{Yuk} + V^{ij}_{mdi} + V^{ij}_{Coul} + V^{ij}_{Sym}$
is the total interaction potential.

3) NN collisions and Particle production.
Collision between two nucleons takes place only if the following condition holds

$$d < |r_i - r_j| \leq \sqrt{\frac{\sigma_{nn}}{\pi}}$$

## Results and Discussions

The ALADIN results are the most complete piece of data available for multifragmentation. In heavy ion collisions energy depositions are reached which covers the range of particle evaporation to multifragmentation emission and the total disassembly of the nuclear matter so called the 'rise and fall' of multifragmentation [3]. The most prominent feature of the multifragment decay is the universality of the fragment and fragment charge correlation. The loss of memory of the entrance channel is an indication that the equilibrium is attained prerior to the fragmentation stage of the reaction. Here we are comparing our results with experimental data of reactions $_{57}La^{124}+_{50}Sn^{124}$, $_{50}Sn^{124}+_{50}Sn^{124}$ and $_{50}Sn^{107}+_{50}Sn^{124}$ at energy 600 MeV/nucleon[3]. In Fig. 1 we have shown IMF's as a function of $Z_{bound}$. The quantity $Z_{bound}$ is defined as sum of all atomic number $Z_i$ of all projectile fragment with $Z_i \geq 2$. We observe that at semi peripheral collisions multiplicity of IMF's shows fall' in multifragmentation emission. It is observed that IMF's shows the agreement with data at low impact parameters but fails at intermediate impact parameters. We shall try to reproduce the result with reduced cross-section in future.

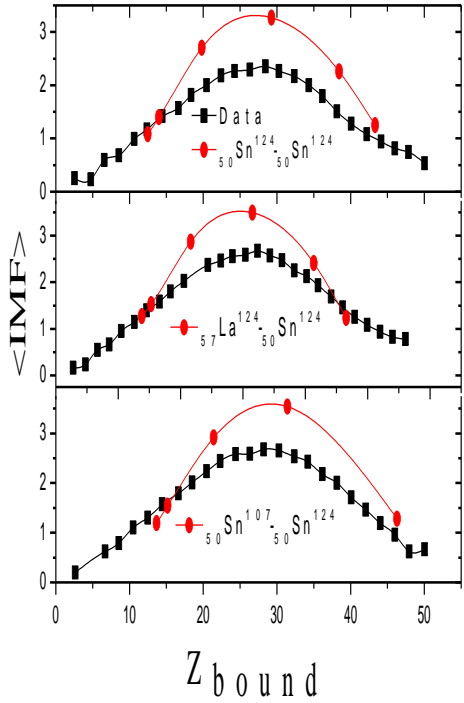

**Fig. 1** Multiplicity of IMF's as a function of $Z_{bound}$

a peak. In case of central collisions, the collisions are so violent that few number of IMF's are observed for peripheral collisions very small portion of target and projectile overlap so again few number of IMF's observed most of the fragments produce as heavy mass fragments (HMF's). In this way we get a clear 'rise and